# Leveraging Unsupervised Learning to Summarize APIs Discussed in Stack Overflow

AmirHossein Naghshzan
École de technologie supérieure
Montreal, Canada
amirhossein.naghshzan.1@ens.etsmtl.ca

Latifa Guerrouj
École de technologie supérieure
Montreal, Canada
latifa.guerrouj@etsmtl.ca

Olga Baysal
Carleton University
Ottawa, Canada
olga.baysal@carleton.ca

*Abstract*—Automated source code summarization is a task that generates summarized information about the purpose, usage, and–or implementation of methods and classes to support understanding of these code entities. Multiple approaches and techniques have been proposed for supervised and unsupervised learning in code summarization, however, they were mostly focused on generating a summary for a piece of code. In addition, very few works have leveraged unofficial documentation.

This paper proposes an automatic and novel approach for summarizing Android API methods discussed in Stack Overflow that we consider as unofficial documentation in this research. Our approach takes the API method's name as an input and generates a natural language summary based on Stack Overflow discussions of that API method. We have conducted a survey that involves 16 Android developers to evaluate the quality of our automatically generated summaries and compare them with the official Android documentation.

Our results demonstrate that while developers find the official documentation more useful in general, the generated summaries are also competitive, in particular for offering implementation details, and can be used as a complementary source for guiding developers in software development and maintenance tasks.

## I. INTRODUCTION

Researchers found *automatic summarization* an interesting topic in the 1950s, especially when Luhn [1] published his paper in this domain. Nowadays, summarization has gained much attention [2], in particular, automatic code summarization that is one of the challenging topics in software engineering [4].

*Automatic code summarization* is a task that automatically generates summaries, in the form of natural language descriptions, to help software developers understand and comprehend source code quickly [3]. In effect, code summaries are the key for a better understanding and comprehension of source code in software development and maintenance [4]–[6]. Additionally, code summaries allow developers to spend less time understanding the code when trying to meet/update software requirements. For all these reasons, automatic source code summarization is one of the engineering tasks that have received attention and still a topic of interest for many researchers nowadays.

In many cases, developers are unaware of the purpose or usage of a code entity. They must examine a large volume of code and–or amount of documentation to grasp the concept related to a code entity. It is therefore interesting to have an automatic approach that can provide him/her with summaries on the purpose, implementation, and–or usage of code entities that are part of their tasks. Consider, for instance, a developer who is trying to fix a bug that was caused by someone else's task. To understand the bug and how to reproduce it, developers must first read all related bug reports and review previous discussions. Also, a developer who wants to implement a function for the first time needs to read all its related documentation to become familiar with the method and understand how to use it. Although developers use official documentation as their main source of information about code entities [16], researchers have shown that official documentation sometimes lacks completeness, insights, and conciseness [17], [18]. As a result, developers may refer to other sources (e.g. Stack Overflow, GitHub) as well to get insights about a code entity, its implementation, usage, and further information that might not be offered by official documentation. These kinds of sources that are presented by the software community are named unofficial documentation.

In this work, we propose an automatic approach to summarize APIs by leveraging unofficial documentation and unsupervised learning. In this study, we use Stack Overflow as a type of unofficial documentation for our investigation. In addition, we focus on extractive code summarization, which extracts the most important sentences from documents, i.e., Stack Overflow posts in our research.

The goal is to guide and assist developers during their software maintenance and evolution tasks, by helping them quickly understand the API methods that are part of their tasks. The results of this research can be transformed in the form of useful tools that can be used by researchers as well as practitioners in practical settings. As we will discuss in the upcoming sections, our preliminary results show that developers are very interested in the outcome of this study as well as the tool support.

We address the following research questions in this work:

**RQ1**: *Can we leverage unofficial documentation and unsupervised learning to summarize APIs?*

**RQ2**: *Are these automatically generated summaries perceived useful by developers?*

**RQ3**: *Can our automatically-generated summaries compete with the descriptions presented by official documentation?*

## II. RELATED WORK

We focus on the most relevant contribution to this research. First, we present works on code summarization then work that leveraged unofficial documentation in the context of the summarization task.

To summarize methods without any comments in source code, Rodeghero *et al.* [19] has conducted a study using eye-tracking and eye-movement based on a Vector Space Model (VSM) and Information Retrievals (IR) to find important keywords and list them according to a gaze time of the expert developer. They found that the method signature was more important than method calls in source code. As mentioned by Abid *et al.* [37], Rodeghero *et al.*'s approach did not consider the method name, data declaration, or another relevant method, outside of a method.

In another work related to generating documentation for Java methods, McBurney *et al.* [28] have proposed a new approach for producing natural language text by using the context around the method to summarize. This approach is a template-based solution that generates a summary based on a natural language model. Unlike this work, we do not leverage source code neither a template-based solution to summarize code entities. However, our work focus is on unofficial documentation and we leverage unsupervised learning to tackle our problem.

To generate a summary for Java classes using only source code, Moreno *et al.* [30] proposed an approach, based on a combination of stereotype identifiers and lexicalization [32] to understand the content and responsibility of classes. The purpose of this approach is to produce text from the content concerning stereotype identification [31]. The results have shown that the proposed approach can be used as a structured natural language summary for Java classes. Moreover, in a more recent study, Moreno *et al.* [29] has developed the MUSE approach to illustrate the concrete use of the method by code sampling. Oppositely to MUSE, in our research, we present multi-line text summaries for each API instead of code samples.

Recently, the concept of code summarization has received more attention with the use of machine learning and artificial neural network (ANN). These approaches help researchers identify the latent relationship between structural source code and natural language descriptions.

Phan *et al.* [27] have proposed a source code summarization approach based on K-Nearest Neighbors (KNN) and SVM neural network. Using a tree-based approach, in their research, has not only improved the classification performance but also the execution time. Unlike this work, we do not summarize code entities in source code but rather those discussed in unofficial documentation.

In a recent article on source code summarization, LeClair *et al.* [33] have suggested a new approach that generates summaries for code even if there is no internal documentation such as the programmer's comments for methods, especially when there is no good naming. This approach is considered as an improvement of the approach by Hu *et al.* [11] that processes the words from source code to represent an AST. Unlike their research work, we used the TextRank algorithm to generate summaries for Android API methods from unofficial documentation, in particular, Stack Overflow.

Ahmad *et al.* [34] mentioned that using RNN models for code summarization has two major limitations. First, they do not model the non-sequential structure of source code and secondly, as source code could be very long, RNN models fail to capture the long-range dependencies between code tokens. So, they proposed a transformer model approach adopted from Vaswani *et al.* [35].

In terms of using unofficial documentation in the context of the summarization task, Rastkar *et al.* [7] proposed an approach to summarize bug reports. Unlike their work, we do not summarize bug reports or any kind of unofficial documentation. However, we summarize APIs discussed in unofficial documentation, specifically the Stack Overflow platform in this investigation.

Uddin *et al.* [12] tried to summarize the opinion and reviews of people about an API in Stack Overflow. Users can benefit from getting ideas about the limitation of each API and other users' opinions about it. The main difference of this work with our research is that we are interested in summarizing APIs in terms of their purpose, implementation/usage rather than people's opinions about them.

In a more recent paper by Uddin *et al.* [13], the authors extended their previous research to cover the usefulness of APIs as well. They added a code block to each summary to show API usage. In contrast, in our research, we do not use code blocks. We are only interested in generating natural language descriptions that summarize APIs and methods. Moreover, another major difference is that in their work each description consists of three parts: title, problem, and solution. It means that their research summarizes an API's problems and possible ways to solve them. However, in our case, we do not focus only on problematic APIs.

Saddler *et al.* [26] conducted a study using eye-tracking methods to summarize API elements in Stack Overflow posts as input sources. To evaluate their approach, the authors invited 30 participants, consisting of students and professionals. Although there was not much difference in the accuracy between the two groups, there was a difference in how each page's content was viewed. Unlike this work, we do not rely on eye-tracking methods, however, we leverage unsupervised learning to generate summaries for methods discussed in unofficial documentation.

In another work, Jiang, Armaly, and McMillan [14] conducted a study to generate short commit messages that summarize software changes based on the difference between two versions of the source code with the Neural Machine Translation (NLT) algorithm. A corpus that is used for training their algorithm is prepared from a GitHub project. In contrast, We used the TextRank algorithm to generate a summary for the usage of Android methods from unofficial documentation. Our research generates summaries automatically and without human intervention.

Iyer *et al.* [36] presented a model (CODE-NN) Recurrent Neural Network (RNN) to generate high-level summaries for C# code snippets and SQL queries. The model is trained based on Stack Overflow posts and the results prove that the model improves the state of the art. Similar to their work, we use Stack Overflow as a source to build our corpus and train our model. However, we are not interested in generating a summary for a code snippet. Our approach takes as input an API method's name and generates a natural language summary for that API method, using Stack Overflow posts in which the API method has been discussed.

Overall, there is a large body of work that has exploited source code to summarize code entities. However, only limited research studies have been conducted on using unofficial documentation. In this work, we investigate whether we can gain more insights from unofficial documentation using unsupervised machine learning. In addition, we explore the perceptions of developers on the automatically produced summaries.

## III. METHODOLOGY

In this section, we describe the main phases of our methodology for summarizing APIs discussed in Stack Overflow (Cf. II): Data Collection, extraction of APIs mentioned in Stack overflow, building a corpus for each API, pre-processing the corpus, summarization (vectorization and applying TextRank), empirical evaluation, data analysis, and generation of the results.

### A. Data Collection:

We used the Stack Overflow platform as a type of unofficial documentation and considered only posts with the "Android" tag. We choose Android since it is one of the top eight most discussed topics on Stack Overflow based on the type of tags assigned to questions[1]. Moreover, the first author is familiar with Android development, making it, therefore, more convenient for us, when understanding the different methods and comparing the results.

We have used Stack Exchange API to extract all Stack Overflow's questions tagged as "Android", from January 2009 till April 2020. As shown in Table I, we have collected 1,266,269 unique Android questions. Then, we have extracted all the answers for the collected questions that resulted in 1,817,874 answers. In total, we have gathered 3,084,143 unique Android posts from Stack Overflow, which formed our dataset.

### B. Extraction of APIs:

The next step is to extract APIs mentioned in Stack Overflow. We have thought about using advanced techniques from the literature such as Baker, a tool presented by Subramanian, Inozemtseva, and Holmes [15] to extract API methods from Stack Overflow. However, since Stack Overflow offers the *Code Snippet* feature since August 25, 2014, we have used it to extract this information. The idea is to parse the body of posts and find available code snippets within sentences. This feature highlights code blocks in a post to make it easier for users to spot and use code. If a user wants to mention a piece of code in their post, they need to put the code between quotations for Stack Overflow to detect it as a code snippet. Such a rule is enforced by Stack Overflow, while the user may forget to put the code inside quotations, other users can edit the post and put the code between quotations. In the pure HTML format of the post, Stack Overflow changes quotation to an HTML tag (`<code>`). By automatic parsing the body of posts and extracting the elements that are surrounded with `<code>` tag, we find code snippets within posts. However, there are two types of code snippets. The first type is usually the name of functions and APIs and the second type is a multi-line code block which is a sample of usage and/or implementation of a function that we refer to as a *code sample*. Figure 2 shows an example of these two code snippets in Stack Overflow.

[1] https://en.wikipedia.org/wiki/Stack_Overflow

TABLE I
DATA EXTRACTED FROM STACK OVERFLOW.

| Posts | Count |
| --- | --- |
| All Questions | 21,165,633 |
| Android Questions | 1,266,269 |
| Answers to Android Questions | 1,817,874 |
| Total of Android Posts | 3,084,143 |

We parsed the posts and extracted all code entities to identify the most discussed methods. In some cases, the code entities are not fully qualified. For example, the third code entity in Figure 2, *android:configChanges="orientation—screenSize"*, is not an API name. To overcome this issue and validate the API names we tried a semi-automatic process to verify code entities. First, we used regex patterns to detect irrelevant code entities. Android method names are referred to as *identifiers*. An identifier is an unlimited-length sequence of letters and digits without any special character except for the underscore (i.e., "_") . Therefore, we kept code entities consist of only letters, digits, underscore, dots and parentheses using this pattern "`^[^a-zA-Z0-9_.()]`". We kept dots and parentheses as the code entities may be fully qualified API names with package names, and in some cases, users write the methods with their parentheses or even arguments, similar to the first code entity shown in Figure 2. However, in many cases, they use file names like `studio.sh`. Therefore, if a code entity matches `^(\/+\w{0,}){0,}\.\w{1,}$`, it is a file, so we removed those that match this regex pattern. And finally, the first author manually checked the remaining code entities to remove the irrelevant ones and those that are not valid Android methods. During the manual verification, 37 code entities were eliminated. They were mostly Android class names such as `Activity` or `TextView`.

After omitting irrelevant items and cleaning the functions, we found that the most occurring method in Android posts is `app.Activity.onCreate()` as reported in Figure 2. This method has been mentioned 21,103 times in answers and 7,933 times in questions. With 29,036 mentions, `app.Activity.onCreate()` appears to be the most popular Android method in Stack Overflow. In Figure 3, we illustrate the most repeated Android methods in Stack Overflow.

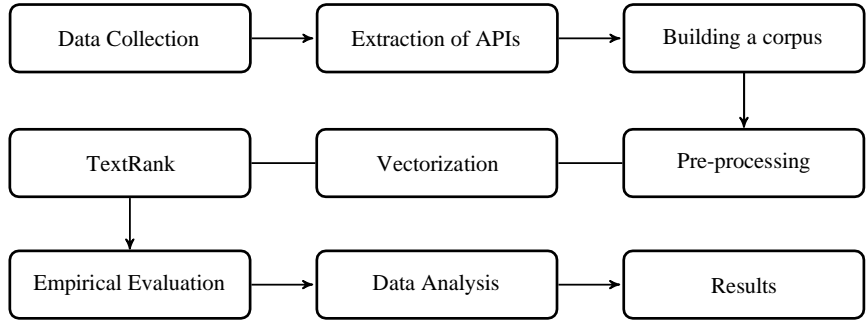

Fig. 1. Main phases of the methodology.

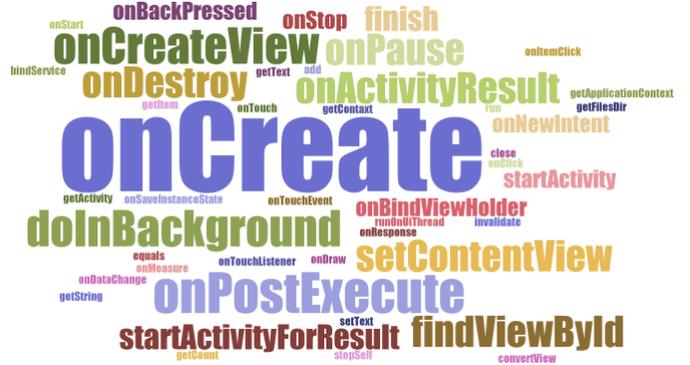

Fig. 2. An example of code snippets in Stack Overflow.

We sorted the methods based on their repetition, removed the deprecated API methods, and selected the 15 most repeated Android methods for the summarizing task. We selected the top 15 APIs because the repetition of other APIs suddenly decreased to a few hundred which shows they are not as popular as the top 15 methods in SO. Table II shows the repetition of these methods in Stack Overflow posts.

*C. Building a Corpus:*

Creating a corpus is one of the most important phases of the summarization task. As we are working on extractive summarization, our corpus should incorporate sentences from Stack Overflow posts. In the following, we explain how to build a corpus for the summarization process.

It is important to mention that we build a corpus for each method since each one has its own context and its corresponding Stack Overflow posts in which it has been discussed. To build a corpus for each method to be summarized, we need to select posts with pertinent information related to a method in question but the challenge is how to choose these posts.

Fig. 3. Most repeated Android methods in Stack Overflow.

TABLE II
TOP ANDROID METHODS BASED ON THE NUMBER OF REPETITION.

| Method with Qualified Name | Question | Answer | Total |
|---|---|---|---|
| app.Activity.**onCreate**() | 7,933 | 21,103 | 29,036 |
| os.AsyncTask.**onPostExecute**() | 1,413 | 5,452 | 6,865 |
| app.Fragment.**onCreateView**() | 1,769 | 4,291 | 6,060 |
| app.Activity.**onActivityResult**() | 2,196 | 3,853 | 6,049 |
| os.AsyncTask.**doInBackground**() | 1,135 | 4,569 | 5,704 |
| app.Activity.**onPause**() | 1,556 | 4,031 | 5,587 |
| app.Activity.**findViewById**() | 804 | 4,182 | 4,986 |
| app.Activity.**onDestroy**() | 1,588 | 3,333 | 4,921 |
| app.Activity.**finish**() | 1,105 | 3,224 | 4,329 |
| app.Activity.**setContentView**() | 627 | 3,562 | 4,189 |
| app.Activity.**onStop**() | 949 | 2,188 | 3,137 |
| app.Activity.**startActivityForResult**() | 779 | 2,325 | 3,104 |
| recyclerview.**onBindViewHolder**() | 845 | 1,973 | 2,818 |
| app.Activity.**startActivity**() | 558 | 2,217 | 2,775 |
| app.Activity.**onBackPressed**() | 652 | 1,906 | 2,558 |
| Total | 23,909 | 68,209 | 92,118 |

For each method, we extracted the SO posts that the method appeared in. However, based on our experience, including Stack Overflow questions in the corpus of a method may result in summaries that have interrogative (i.e., question) sentences instead of common declarative sentences (i.e., containing facts, explanations, information, etc.). Therefore, we leveraged only Stack Overflow answers for building our corpus plus titles. We believe titles are one-line summaries of questions that contain important information about the posts. Furthermore, not all SO answers are useful and some of them can contain incorrect information that may mislead developers instead of guiding them. Adding such answers to the corpus (of a method in our approach) may lead to including incorrect information in our summaries. To minimize this risk and improve the quality of our corpus, we have selected answers with a score higher than the average score for all Android answers in our dataset, which is 2.87. Thus, we selected SO answers with a score equal to or higher than 3 (scores are integer numbers). Once the SO posts have been selected, we have considered their bodies to build our corpus. However, considering all the sentences that exist in a body of a Stack Overflow post maybe not that relevant since irrelevant sentences may exist in SO posts. One way to overcome this issue is to consider relevant sentences (criteria 1 and 2) in the context of the summarization task, as well as *proximity* (criteria 3 and 4) as in previous works [8], [9] other researchers suggested:

1) The first sentence of each post.
2) The main sentence containing the method.
3) A sentence before the main sentence.
4) A sentence after the main sentence.

We have used the standard Natural Language Toolkit (NLTK) sentences Tokenizer to split our text into sentences[2]. By following the above steps, we have created a corpus for each API method in our dataset.

*D. Pre-Processing:*

Once a corpus has been built for each API method, we have applied the following pre-processing steps to prepare the corpus for the summarization algorithm.

- Removing punctuation, numbers, HTML tags, and special characters from sentences.
- Removing stop words from the sentences using Natural Language Toolkit (NLTK).
- Lemmatization has been applied to get the root of words so that the weight of the words can be calculated correctly.
- Duplicate sentences have been excluded from the corpus.
- Text Vectorization has been applied to transform our documents into vector representations such that we can apply numeric machine learning.

There are many ways to transform a text into a vector such as Bag Of Word (BOW), TF-IDF, as well as deep learning models such as word-embeddings. We have used word embeddings to represent the data for summarization. Word embeddings are a family of natural language processing techniques aiming at mapping semantic meaning into a geometric space. Word embedding can be useful for most NLP problems, especially if there is not a lot of training data. And that is the reason, we have used it since the number of posts varies for each method. Some have a high number of posts in which they have been discussed and vice-versa. For the word embedding technique, we have used Stack Overflow's pre-trained word embeddings proposed by Efstathiou *et al.* [10]. They have released a word2vec model trained over 15GB of textual data from Stack Overflow posts.

[2]https://www.nltk.org/

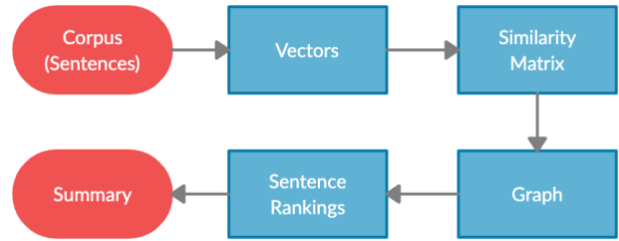

Fig. 4. Overview of the TextRank algorithm.

*E. Summarization:*

To summarize methods discussed in Stack Overflow, we have used unsupervised learning, specifically the TextRank algorithm, which is appropriate for the summarization task in an extractive way [25]. TextRank algorithm is a graph-based unsupervised extractive summarization technique introduced by Mihalcea and Tarau in their highly-cited paper [25]. TextRank algorithm was originally inspired by Google's famous PageRank algorithm [20] to find representative key phrases in a given dataset [12]. According to Mihalcea and Tarau [25], TextRank has been proven to generate high-quality text summaries. This algorithm consists of two major steps: text similarity and PageRank algorithm.

- To estimate the similarity between the sentences, the cosine similarity method has been applied. As Baoli *et al.* [38] have explained, the cosine similarity method is applied to investigate the similarity between the two vectors. This method can be used for text classification and text summarization also. In this approach, the similarity between sentences has been calculated based on the cosine similarity.
- PageRank is an algorithm used by Google Search to rank websites in their search engine results. To find the most appropriate answers to a given question, PageRank is combined with several other factors such as standard IR metrics, proximity, and anchor text. [23]. It is a way of measuring the importance of website pages [20], [21]. The basic idea is that if page A contains a connection to page B, the author of page A is implicitly endorsing page B. [22]. In the TextRank algorithm, PageRank calculates the importance of each sentence and gives a rank to each sentence. The final summary will be generated based on the top-ranked sentences.

As shown in Figure 4, the main steps of the TextRank algorithm are as follows:

- As described before, the first step is to split SO posts into individual sentences and select relevant sentences

- from each post and combine all those sentences to form a corpus.
- In the next step, vector representation (word embeddings) is applied for each sentence.
- Similarities between sentence vectors are then calculated and stored in a matrix using cosine similarity
- The similarity matrix is then converted into a graph, with sentences as vertices and similarity scores as edges, for sentence rank computation.
- Finally, top-ranked sentences to be included in a summary are chosen based on a threshold to form our final summary.

The outcome of the TextRank algorithm is a list of sentences ranked based on their importance for each method. The last step is to combine these sentences to form a summary, however, the length of summaries is important in the summarization task. That is the reason why we have used a threshold to select top-ranked sentences. In effect, if a summary is too long, it would be too time-consuming and overwhelming for humans. And if it is too short, it may not capture all the necessary information. To determine a reasonable length for the automatically generated summaries, we have selected the top three sentences for each method to generate the final summary. This threshold is based on our experience to fulfill mentioned concerns. Table III shows examples of summaries for five randomly selected methods from our data.

## IV. Empirical Evaluation

Since our long-term goal is to provide developers with tool support to help them quickly understand APIs that are part of their tasks, we have conducted an empirical investigation to gain insights on the perceptions of developers on the quality of the automatically generated summaries and the usefulness of our approach.

We follow the Basili framework [24] to describe our study.

### A. Definition and Planning of the Study

The *goal* of our investigation is to evaluate the quality and usefulness of the automatically-generated summaries. The *purpose* of this research work is to help developers quickly understand APIs and methods that are part of their software engineering tasks.

The *quality focus* of our study is measured in terms of the usefulness, coherence, accuracy, and relevance of the automatically-generated summaries.

The *research perspective* is of researchers and Android developers interested in having automatic approaches and tools that generate automatic summaries for APIs and code entities.

The *context* consists of our dataset of Android Stack Overflow posts, gathered between January 2009 to April 2020.

### B. Research Questions

We have addressed the following main research questions during our empirical evaluation:

**RQ1: Can we leverage unofficial documentation and unsupervised learning to summarize APIs?**

It is important to generate high-quality summaries for APIs. Quality of summaries is measured at this level in terms of coherence, accuracy, and length.

**RQ2: Are these automatically generated summaries perceived useful by developers?**
The main goal of these summaries is to help developers save time and be efficient. Thus, it is important to know if we could satisfy developers.

**RQ3: Can our automatically-generated summaries compete with the descriptions presented by official documentation?**
We are interested to compare our summaries with official descriptions to check if the generated summaries are as good as the official's in terms of usefulness.

### C. Study Design

We have used a completely randomized design [40] when conducting our survey with Android professional developers because our participants mostly had the same level of Android development skills. Therefore, we could not generate blocks based on their expertise and familiarity with Android development. With a completely randomized design, participants are randomly assigned to treatments.

Regarding the tasks, participants were asked to perform three tasks on a set of 3 Android API methods. We assigned only three APIs to each participant to prevent confusion and fatigue. The tasks consist of the following:

- For each task, for each Android API method, we asked each participant to write a summary for the Android API method in question. For each API method, we provided participants with a Stack Overflow link. This link allowed participants who are unfamiliar with a specific API method to get some insights about that method before writing the summary. Our goal for this task was to help enhance the quality of automatically-generated summaries.
- Then, we have presented our automatically-generated summary for that Android API method and asked participants to evaluate the summary based on a set of criteria: the coherence of a summary, its accuracy, its length, and relevance. These criteria are the widely-used ones in the software engineering literature when evaluating summaries by conducting studies with developers [14], [31].
- For the last task, we have presented the official description that exists in the official Android API documentation for the API method in question and asked the participants to compare the official description with the generated summary in terms of usefulness and completeness. Descriptions are summaries presented by the official Android Developer website[3]. These descriptions could be a short one-line summary or for example a page depending on the complexity of methods.

[3]https://developer.android.com

TABLE III
EXAMPLES OF GENERATED SUMMARIES.

| Method | Summary |
|---|---|
| app.Activity.**onCreate()** | When the app is launched, the first thing that's going to run is onCreate() in this case, onCreate() has a method that inflates the view of your activity, that method is called setContentView(). When you pass data from one activity to another using a Bundle, the data is received inside the onCreate() method of the second activity, not insideonActivityResult() unless you've specifically implemented that. Inside your Activity instance's onCreate() method you need to first find your Button by its id using findViewById() and then set an OnClickListenerfor your button and implement the onClick() method so that it starts your new Activity. |
| app.Activity.**startActivity()** | if this method is being called from outside of an Activity Context, then the Intent must include the FLAG_ACTIVITY_NEW_TASK launch flag. What you can do is to simply start the activity using startActivity() and have the called activity call startActivity() to return to your activity, sending back data as necessary as extras in the Intent it uses. You still only need to call startActivity() to launch a different Activity. |
| app.Activity.**onActivityResult()** | To get the returning data from your second activity in your first activity, just override the onActivityResult() method and use the intent to get the data. In the second activity perform whatever validation you need to do (this could also be done in the first activity by passing the data via an Intent) and call setResult(int resultCode, Intent data) (documentation) and then finish(); from the second activity. From your activity1 start activity for result, and from activity2 use the setResult(int resultCode, Intent data) method with the data you want your activity1 to get back, and call finish() (it will get back to onActivityResult() in the same state activity1 was before). |
| os.AsyncTask.**doInBackground()** | To get result back in Main Thread you will need to use AsyncTask.get() method which make UI thread wait until execution of doInBackground is not completed but get() method call freeze the Main UI thread until doInBackground computation is not complete. You will need to return some data (probably contactList) from your doInBackground() method, and then move the offending code to the onPostExecute() method, which is run on the UI thread. You most likely don't want to directly instantiate a Handler at all... whatever data your doInBackground() implementation returns will be passed to onPostExecute() which runs on the UI thread. |
| app.Activity.**finish()** | Make sure you are calling finish() in onBackPressed() in Activity B; which indicates that you no longer need this Activity(Activity B) and can resume last activity which was paused/stopped and is in background stack Basically, use setResult before finishing an activity, to set an exit code of your choice, and if your parent activity receives that exit code, you set it in that activity, and finish that one, etc... finish() is only called by the system when the user presses the BACK button from your Activity, although it is often called directly by applications to leave an Activity and return to the previous one. |

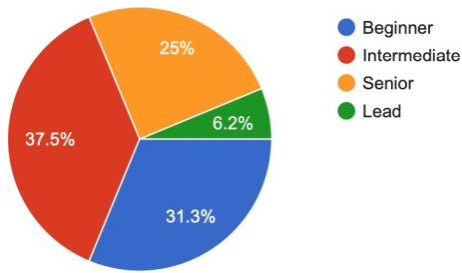

Fig. 5. Participants' Android development skills.

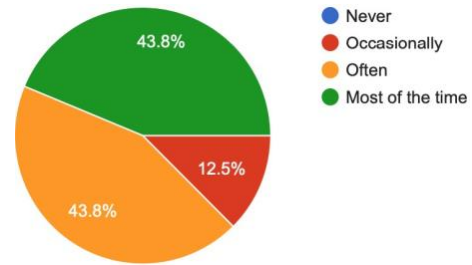

Fig. 6. How often participants use Stack Overflow.

- As the last question, we wanted to know if our summary can be used as a complementary source for official Android documentation.

*D. Participants*

After having the approval of the ethical committee of our institution, we have contacted 28 Android developers by email and asked them to participate in our questionnaire. Among these 28 developers, 16 developers replied and agreed to participate in our survey. Participants were mostly professional developers, as well as students who had prior industrial experience with Android development. Our participants include both male and female professional developers with different levels of Android development skills. We have selected participants with industrial experience to make sure our results are valid from an industrial point of view and can be used in long term by professional developers in practical settings.

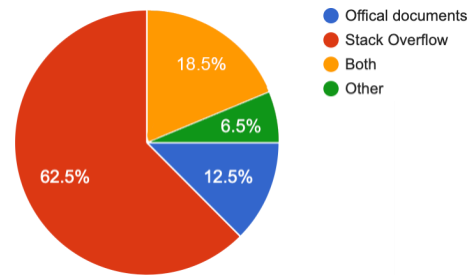

Fig. 7. Use of Stack Overflow vs. official documentation.

According to the pre-questionnaire, 62.5% of participants are male, while 37.5% are female. As shown in Figure 5, most of the participants have intermediate expertise in Android development and all of them use Stack Overflow for development purposes (Cf. Figure 6). Most of the participants (56.2%) are developers currently working in companies and the rest (43.8%) are students who also have work experience in the industry. Moreover, we asked participants about the information sources that they refer to during their development activities. As stated in Figure 7, most developers (62.5%) use Stack Overflow as the main source of gaining information about API methods. 18.5% use both the Stack Overflow and the official documentation, while only 12.5% of participants use official documentation as the main source of information. These numbers demonstrate that Stack Overflow is a more preferred choice among our participants making it a valuable and rich source of information.

*E. Questionnaire*

As previously discussed in Table II, we have selected the 15 most repeated Android methods to generate summaries. We presented the automatically-generated summaries to the participants to evaluate them and compare them with the official Android descriptions. To mitigate the *courtesy bias* of the study, we did not indicate which summary was extracted from official documentation and which one was automatically-generated by our approach. For each API method, the summaries were presented to the participants in the randomized order. In some cases, participants first evaluated the generated summaries and then the descriptions from the official documentation. In other cases, they were asked to evaluate the official descriptions first. The questionnaire includes questions about the accuracy of the summaries, their coherence, the length of the summaries, the usefulness of summaries, as well as participants' opinions/feedback to further improve our automatically-produced summaries. Those who accepted to be part of our study received instruction about the questionnaire to know what are the steps, how to answer the questions, and also have an idea about general concepts like coherence, accuracy, and usefulness. Table IV shows summary of questions presented to participants.

## V. RESULTS AND DISCUSSION

We have evaluated a total of 3,084,143 unique Stack Overflow posts and summarized the top 15 most popular Android APIs. In the following, we present the most important findings of our survey:

- As it can be noticed from Figure 8, all developers involved in this study (100%) agree that the length of summaries is appropriate. Although we used a fixed number of sentences as a summary for each method (3 sentences), the length of sentences depends on the complexity of the methods. Usually, for simple methods, the sentences are shorter in comparison to complex methods.

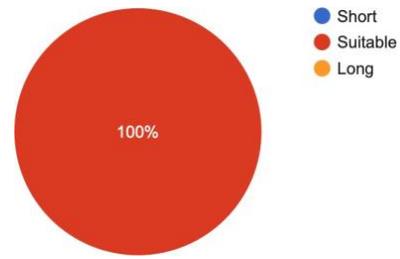

Fig. 8. Developers' perception on the length of summaries.

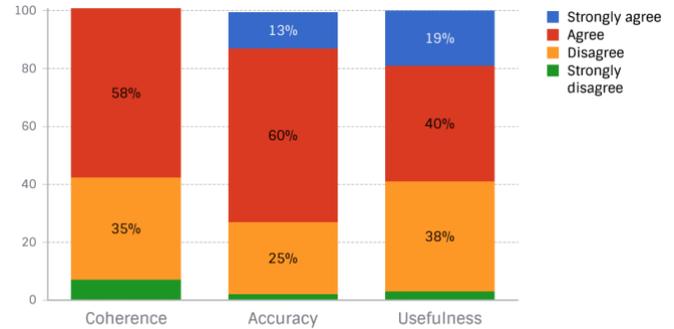

Fig. 9. Developers' satisfaction with the quality of generated summaries.

- We asked our participants about the quality of summaries, and as mentioned earlier it was measured using three metrics as presented in Figure 9. In the case of coherence, as can be noticed from the chart, about half of the participants (58%) believe that the automatically-generated summaries are coherent. This surprising result may be very likely since, in this approach, summarization is done in an extractive fashion, which means that the relevant sentences are selected, (in this case, based on unsupervised learning), from a corpus that has been built using different SO posts, and then simply combined to form our summaries, which may have affected the coherence of the produced summaries. We are aware of that and plan to tackle this challenge as part of our future work.

- As shown in Figure 9, most of the participants (73%) found that our summaries include accurate information about Android methods. The remaining percentage can be explained by the fact that, in some cases, the answers to a Stack Overflow question are wrong or contain misleading information. Dealing with such posts may result in generating inaccurate information that can be parts of the generated summaries. We mitigated this thread by selecting answers with a score of 3 and higher.

- The results also show that 59% of participants (Figure 9) believe that automatically-generated summaries contain important and necessary information about Android methods. Not surprisingly, the automatically-generated summaries do not contain all information about Android methods since they rely on one source only (Stack Overflow posts only) but developers could use them to get important information. We are aware of this challenge and plan to augment our automatic summaries by considering other types of unofficial documentation in our future work.

TABLE IV
SUMMARY OF QUESTIONS.

| Question ID | Question |
|---|---|
| Question 1 | Is the summary coherent? |
| Question 2 | Is the summary accurate? |
| Question 3 | How did you find the length of summaries? |
| Question 4 | Does the summary contain all information about the method? |
| Question 5 | Does the summary contains only important information about the method? |
| Question 6 | Does the summary contain information that helps understand how to implement the method? |
| Question 7 | Does the summary contains information that help understand how to use the method? |
| Question 8 | Does the summary contains information that help understand how to implement the method? |
| Question 9 | Can the summaries help developers to reduce their effort of searching relevant information? |
| Question 10 | Can the summaries help developers to reduce the time of development? |
| Question 11 | Can the summaries be helpful for developers in any steps of development? |
| Question 12 | Which summary do you find more useful? |
| Question 13 | Which summary can help you better understand the usage of the method? |
| Question 14 | Which summary can help you better understand the implementation of the method? |
| Question 15 | Do you think the summaries can be used as a complementary source for each other? |
| Question 16 | Is it helpful to have a summarizer in within your IDE to get information about methods? |

- When it comes to comparing the automatically-generated summaries with official Android documentation (Figure 10), there are not many differences between the two: almost the same proportions have been obtained for both, except for a slightly smaller difference in terms of whether it is for implementation versus usage. In effect, it seems that participants are more in favor of using the automatically-generated summaries when implementing Android methods, while they are more towards official documentation when they use a method. Although this result may seem surprising, the difference is very small. Further studies are required to gain more insights into this.

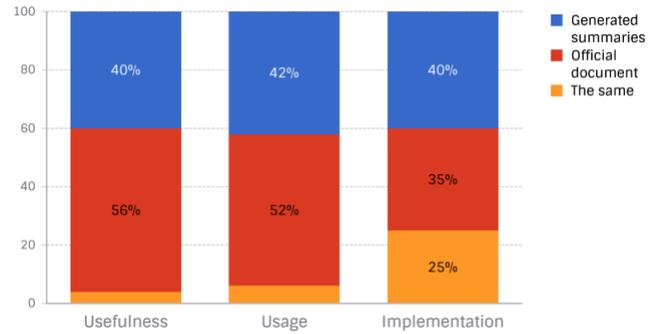

Fig. 10. Which one is better – generated summaries or official documentation?

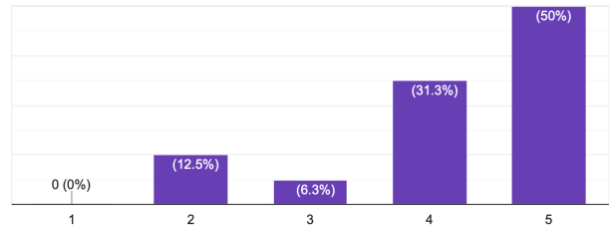

Fig. 11. Would it be helpful to have a summarizer plugin within IDE?

- With a rate of 4.1 out of 5 (Figure 11), participants agreed on the fact that it would be helpful to have an integrated plugin to show our automatically-generated summaries. This feedback gives us a certain encouragement that the produced summaries could be further enhanced and our approach can be implemented to use as tool support in a practical setting by software developers.
- With the rate of 3.6 out of 5 (Figure 12), participants agree that it is a good idea to augment the automatically-generated summaries with official documentation. As mentioned before, researchers have shown that official documentation is sometimes incomplete, not concise, or may lack insights [17]. Therefore, this result makes sense and could be taken into account in future work.

Overall, these findings show that:

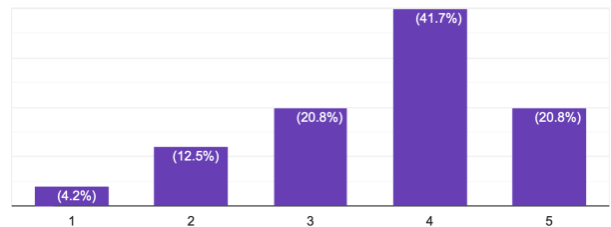

Fig. 12. Can summaries be used as a complementary source?

For **RQ1** to check whether we can use unofficial documentation and unsupervised learning for summarization propose, we can generate automatic summaries for methods discussed in Stack Overflow. The summaries have been perceived as accurate and relevant by Android developers.

Although we could achieve promising results, our approach failed to generate summaries for some methods. The main reason for these failures is the lack of data. We tried to generate summaries for all methods that have been part of our initial dataset including unpopular methods. As expected, we ended having only tens of sentences for such methods. And since machine learning requires a reasonable amount of data to provide reasonable results, we have dealt with popular methods only for this initial study. However, we have reflected on this problem during our process. And it can be solved by considering other types of unofficial documentation or other sources to enrich our dataset. To wrap up, we believe that our findings still convey an interesting message to the community interested in code summarization: it is possible to use unofficial documentation to generate code summaries. We will continue to improve our results and the quality of the produced summaries.

Regarding **RQ2** to evaluate our generated summaries by developers, we have conducted a survey, in which 16 Android developers have kindly agreed to participate, and we have asked them to evaluate the automatically-generated summaries to determine if our automatic code summarizes can be useful for them during the software development process or not. The results have shown that developers perceived the automatically-generated summaries as practical.

Finally, for what concerns **RQ3** about comparing our generated summaries with official descriptions, we have asked the participants to read the official descriptions and compare them with the automatically-generated summaries. Overall, the findings have shown no many differences between the two. Additionally, participants agreed that our generated summaries could be used as a complementary source for official documentation. These results inspire us to continue our research and further improve the quality of our automatically-produced summaries so that we can leverage them as tool support for developers and–or use them to enrich official descriptions.

## VI. THREATS TO VALIDITY

Despite the obtained results, there are threats to validity related to our empirical study. We now discuss the most prevalent ones.

### A. Internal Validity

We have tried to keep our survey simple with a reasonable length to prevent confusion and fatigue. However, we are aware that the results could be affected by other external factors like environmental factors, etc. In addition, we have studied Stack Overflow posts from 2009 to 2020. During this time different versions of APIs were released. While we tracked the changes for the examined API methods over the years, a new version of an API may be in contrast with the previous posts of Stack Overflow. We mitigated such a threat by detecting and deleting deprecated methods from our corpus.

### B. External Validity

Our approach is limited to the Android mobile platform and APIs discussed in Android Stack Overflow posts. Extending our approach to other programming languages and domains would be desirable. Another threat is related to the restricted number of APIs. In our study, we have selected the top 15 most repeated APIs from these Android packages: *android.app, android.widget, android.os*, and *androidx.recyclerview*. More methods need to be examined to generalize the obtained results. Another limitation is related to the quality of data needed to generate API summaries. We considered only the most popular methods to avoid generating low-quality summaries for methods with poor Stack Overflow coverage.

### C. Conclusion Validity

We attempted to provide all the necessary details to replicate our study. The details of our survey and responses are provided in our online appendix [39]. However, using the results for other domains requires a detailed analysis of that domain to overcome the possible uncertainties.

## VII. CONCLUSION AND FUTURE WORK

In this study, we have presented a novel code summarization approach for methods based on unofficial documentation and unsupervised learning. We used Stack Overflow Android posts as our dataset and applied the TextRank algorithm as our main technique for summarization. For the pre-processing step, we tried the word-embedding technique to vectorize sentences. The generated summaries were evaluated by 16 professional developers. We found that our automatically-generated summaries could be useful for developers during software development. Additionally, the produced summaries are almost as useful as official documentation when it comes to understanding the usage and implementation of Android methods. Moreover, participants agree that the generated summaries can be used as a complementary source for official documentation.

As our future extension, we are interested in comparing this approach with existing state-of-the-art techniques. Currently, we only validated the feasibility of our approach with professional developers. Moreover, we aim to improve the quality of our summaries. The results we have so far are promising but more efforts are required to make our automatic summaries more insightful. For this purpose, we plan to combine other types of unofficial documentation such as GitHub, bug reports, etc. Additionally, we intend to extend our data from Android API methods to other domains and to improve our summarization technique by implementing deep learning algorithms.

## VIII. ACKNOWLEDGEMENT

We would like to thank all the 16 Android developers who kindly accepted to contribute to our evaluation survey even given the hectic times caused by COVID-19. Your input and feedback are highly appreciated.

## REFERENCES

[1] Luhn, H. P. (1958). The automatic creation of literature abstracts. IBM Journal of Research and Development, 2(8), 159–165. doi:10.1147/rd.22.0159

[2] O. Tas and F. Kiyani, "A SURVEY AUTOMATIC TEXT SUMMARIZATION", PressAcademia Procedia, vol. 5, no. 1, pp. 205-213, Jun. 2017, doi:10.17261/Pressacademia.2017.591

[3] Yao Wan, Zhou Zhao, Min Yang, Guandong Xu, Haochao Ying, Jian Wu, and Philip S Yu. 2018. Improving automatic source code summarization via deep reinforcement learning. In Proceedings of the 33rd ACM/IEEE International Conference on Automated Software Engineering. ACM, 397–407

[4] Y. Zhu and M. Pan, "Automatic code summarization: A systematic literature review," arXiv Preprint, 2019. [Online]. Available: https://arxiv.org/abs/1909.04352

[5] Miltiadis Allamanis, Earl T Barr, Premkumar Devanbu, and Charles Sutton. A survey of machine learning for big code and naturalness. arXiv preprint arXiv:1709.06182, 2017.

[6] Chen, H.; Le, T.H.M.; Babar, M.A. Deep Learning for Source Code Modeling and Generation: Models, Applications and Challenges. ACM Comput. Surv. (CSUR) 2020, 53, 1–38.

[7] S. Rastkar, G. C. Murphy, and G. Murray. Automatic summarization of bug reports. IEEE Trans. Software Eng, 40(4):366–380, 2014. 5. A. Bacchelli, M. Lanza, and R. Robbes. Linking e-mails and source code artifacts. In 32nd ACM/IEEE International Conference on Software Engineering, pages 375–384, 2010.

[8] L. Guerrouj, D. Bourque, P. C. Rigby, "Leveraging unofficial documentation to summarize classes and methods in context", Proceedings of the 37th International Conference on Software Engineering (ICSE), vol. 2, pp. 639-642, 2015.

[9] Barthélémy Dagenais and Martin P. Robillard. 2012. Recovering traceability links between an API and its learning resources. In Proceedings of the 34th International Conference on Software Engineering (ICSE). IEEE Press, 47–57.

[10] Efstathiou, V., Chatzilenas, C., Spinellis, D., 2018. "Word Embeddings for the Software Engineering Domain". In Proceedings of the 15th International Conference on Mining Software Repositories. ACM.

[11] Hu, Xing, Ge Li, Xin Xia, David Lo, and Zhi Jin, 2018. "Deep Code Comment Generation.". Proceedings of the 26th Conference on Program Comprehension. ACM.

[12] G. Uddin and F. Khomh, "Automatic summarization of API reviews," 2017 32nd IEEE/ACM International Conference on Automated Software Engineering (ASE), Urbana, IL, 2017, pp. 159-170, doi: 10.1109/ASE.2017.8115629.

[13] Gias Uddin, Foutse Khomh, Chanchal K Roy, Mining API usage scenarios from stack overflow, Information and Software Technology, Volume 122, 2020, 106277, ISSN 0950-5849.

[14] Jiang, S., Armaly, A., McMillan, C. (2017b). Automatically generating commit messages from diffs using neural machine translation. 2017 32nd IEEE/ACM International Conference on Automated Software Engineering (ASE), pp. 135–146.

[15] Siddharth Subramanian, Laura Inozemtseva, and Reid Holmes. 2014. Live API documentation. In Proceedings of the 36th International Conference on Software Engineering (ICSE). Association for Computing Machinery, New York, NY, USA, 643–652. DOI:https://doi.org/10.1145/2568225.2568313

[16] Ponzanelli, L., Bavota, G., Di Penta, M. et al. Prompter. Empir Software Eng 21, 2190–2231 (2016). https://doi.org/10.1007/s10664-015-9397-1

[17] G. Uddin and M. P. Robillard, "How API Documentation Fails," in IEEE Software, vol. 32, no. 4, pp. 68-75, July-Aug. 2015, doi: 10.1109/MS.2014.80.

[18] M. P. Robillard, "What Makes APIs Hard to Learn? Answers from Developers," in IEEE Software, vol. 26, no. 6, pp. 27-34, Nov.-Dec. 2009, doi: 10.1109/MS.2009.193.

[19] Rodeghero, P., C. Liu, P. W. McBurney, and C. McMillan., 2015, "An Eye-Tracking Study of Java Programmers and Application to Source Code Summarization", IEEE Transactions on Software Engineering 41, no. 11.

[20] Page, L., Brin, S., Motwani, R. et al.. (1998). The PageRank citation ranking: bring order to the Web. Proceedings of the 7th International World Wide Web Conference, Brisbane, Australia (pp.161-172)

[21] Sergey Brin, Lawrence Page, The anatomy of a large-scale hypertextual Web search engine, Computer Networks and ISDN Systems, Volume 30, Issues 1–7, 1998, Pages 107-117, ISSN 0169-7552, https://doi.org/10.1016/S0169-7552(98)00110-X

[22] Haveliwala, T. (1999) Efficient Computation of PageRank. Technical Report. Stanford.

[23] Monica Bianchini, Marco Gori, and Franco Scarselli. 2005. Inside PageRank. ACM Trans. Internet Technol. 5, 1 (February 2005), 92–128. DOI:https://doi.org/10.1145/1052934.1052938

[24] Basili VR, Selby RW, Hutchens DH (1986) Experimentation in Software Engineering. IEEE Trans Softw Eng SE12(7):733–7

[25] Mihalcea, R., Tarau, P. (2004). TextRank: bringing order into texts. Proceedings of Conference on Empirical Methods in Natural Language Processing 2004 (pp.404-411), Association for Computational Linguistics.

[26] Saddler, Jonathan A., Cole S. Peterson, Sanjana Sama, Shruthi Nagaraj, Olga Baysal, Latifa Guerrouj, and Bonita Sharif. "Studying Developer Reading Behavior on Stack Overflow during API Summarization Tasks." In 2020 IEEE 27th International Conference on Software Analysis, Evolution and Reengineering (SANER), 195–205.

[27] Phan, Anh Viet, Phuong Ngoc Chau, Minh Le Nguyen, and Lam Thu Bui, 2018. "Automatically Classifying Source Code Using Tree-Based Approaches.", Data and Knowledge Engineering, Special Issue on Knowledge and Systems Engineering", (KSE 2016).

[28] McBurney, P. W., and C. McMillan., February 2016. "Automatic Source Code Summarization of Context for Java Methods.", IEEE Transactions on Software Engineering 42.

[29] Moreno, Laura, Gabriele Bavota, Massimiliano Di Penta, Rocco Oliveto, and Andrian Marcus., 2015, "How Can I Use This Method?", In Proceedings of the 37th International Conference on Software Engineering, ICSE '15. Piscataway, NJ, USA: IEEE Press.

[30] Moreno, Laura, and Andrian Marcus. 2012. "JStereoCode: Automatically Identifying Method and Class Stereotypes in Java Code.", In Proceedings of the 27th IEEE/ACM International Conference on Automated Software Engineering, ASE 2012. New York, NY, USA: ACM.

[31] Moreno, Laura, Jairo Aponte, Giriprasad Sridhara, Andrian Marcus, Lori Pollock, and K. Vijay-Shanker. 2013. "Automatic Generation of Natural Language Summaries for Java Classes.", In 2013 21st International Conference on Program Comprehension (ICPC).

[32] Sridhara, Giriprasad, Emily Hill, Divya Muppaneni, Lori Pollock, and K. Vijay-Shanker., 2010. "Towards Automatically Generating Summary Comments for Java Methods.", In Proceedings of the IEEE/ACM International Conference on Automated Software Engineering.

[33] A. LeClair, S. Jiang and C. McMillan, "A Neural Model for Generating Natural Language Summaries of Program Subroutines," 2019 IEEE/ACM 41st International Conference on Software Engineering (ICSE), 2019, pp. 795-806, doi: 10.1109/ICSE.2019.00087.

[34] W. U. Ahmad, S. Chakraborty, B. Ray, and K.-W. Chang, "A transformer-based approach for source code summarization," arXiv preprint arXiv:2005.00653, 2020.

[35] Ashish Vaswani, Noam Shazeer, Niki Parmar, Jakob Uszkoreit, Llion Jones, Aidan N Gomez, Łukasz Kaiser, and Illia Polosukhin. 2017. Attention is all you need. In Advances in Neural Information Processing Systems 30, pages 5998–6008. Curran Associates, Inc.

[36] Iyer, Srinivasan, Ioannis Konstas, Alvin Cheung, and Luke Zettlemoyer, 2016. "Summarizing Source Code Using a Neural Attention Model.", In Proceedings of the 54th Annual Meeting of the Association for Computational Linguistics.

[37] Abid, Nahla J., Bonita Sharif, Natalia Dragan, Hend Alrasheed, and Jonathan I. Maletic., 2019. "Developer Reading Behavior While Summarizing Java Methods: Size and Context Matters.", In Proceedings of the 41st International Conference on Software Engineering, IEEE Press.

[38] Li, Baoli, and Liping Han, 2013. "Distance Weighted Cosine Similarity Measure for Text Classification." In Intelligent Data Engineering and Automated Learning – IDEAL 2013, edited by Hujun Yin, Ke Tang, Yang Gao, Frank Klawonn, Minho Lee, Thomas Weise, Bin Li, and Xin Yao, 611–18. Lecture Notes in Computer Science. Berlin, Heidelberg: Springer.

[39] Leveraging Unsupervised Learning to Summarize APIs Discussed in Stack Overflow, https://github.com/scam2021-so/SCAM2021, 29 Jun 2021.

[40] Claes Wohlin, Per Runeson, Martin Hst, Magnus C. Ohlsson, Bjrn Regnell, and Anders Wessln. 2012. Experimentation in Software Engineering. Springer Publishing Company, Incorporated.